\documentclass{elsart}

\usepackage{graphicx}
\usepackage{amssymb}

\begin{document}

\begin{frontmatter}



\title{Spectroscopy of Medium to Heavy Single $\Lambda$-Hypernuclei}


\author{Jeff Mcintire}

\address{Physics Department and Nuclear Theory Center, Indiana University, Indiana 47408}

\begin{abstract}
We develop a method for calculating the doublet splittings of select ground-state single $\Lambda$-hypernuclei. 
This hypernuclear spectroscopy is conducted by supplementing the self-consistent single-particle equations with an 
effective interaction, which follows directly from the underlying lagrangian, to simulate the residual particle-hole 
interaction. Our previous investigation, performed using only the leading-order contributions to the particle-hole interaction, was 
inadequate. In the present work, this method is improved upon by increasing the level of truncation in the residual 
interaction to include gradient couplings to the neutral vector meson, and thereby incorporating the tensor force
into the calculation (which is known to play a crucial role in these systems). As a result, we obtain a realistic
description of the effect of the tensor couplings on the doublet orderings and splittings. 
\end{abstract}

\end{frontmatter}

% main text
\section{Introduction}

In the Kohn--Sham framework, the nuclear many-body system is reduced to a set of single-particle 
equations with classical fields \cite{ref:Ko99,ref:Se97,ref:Fu97}. This framework allows one to reproduce the exact 
ground-state energy, scalar and vector densities, and chemical potential, provided that the
mean-field energy functional is accurately calibrated. In some cases, however, the single-particle 
levels are actually weighted averages of multiple states \cite{ref:Dr90}. To illustrate this point, 
consider the ground-state of ${}^{32}_{15}$P${}_{17}$, which has the configuration
$(2{\mathrm s}_{1/2})_{\mathrm p} (1{\mathrm d}_{3/2})_{\mathrm n}$. The angular momenta of the valence 
proton and neutron couple; therefore, the calculated Kohn--Sham ground-state is actually a doublet.
To determine the true level orderings and splittings, one can supplement the Kohn--Sham equations with 
a Tamm-Dancoff Approximation (TDA) analysis of the particle-hole states. The particle-hole matrix elements are sums of two-body 
Dirac matrix elements, and the particle-hole interaction is determined by the underlying effective field theory.
If retardation is neglected, the interaction is given by (Yukawa) meson-exchange potentials (with 
appropriate spin-isospin operators). This approach has been used to study ordinary nuclei \cite{ref:Fu85} and
single $\Lambda$-hypernuclei \cite{ref:Mc04}. The case of single $\Lambda$-hypernuclei is particularly
interesting as no single isovector coupling is allowed, and there are no exchange contributions, since the 
$\Lambda$ and the nucleon are distinguishable. This will allow us to focus on isoscalar exchange in the 
effective interaction. 

There are no free parameters in this TDA analysis. As a result, there are only three possible ways 
to adjust this method: vary the level of truncation in the underlying lagrangian, introduce additional
degrees of freedom, or include higher-order contributions in the particle-hole interaction. It is of interest 
to investigate the effect of these modifications on the accuracy of this approach, particularly the inclusion of
higher-order terms involving gradient couplings. We have applied this specific improvement to calculations of ground-state, 
$\Lambda$-particle--nucleon-hole splittings in single $\Lambda$-hypernuclei, such as ${}^{16}_{\Lambda}$O.
The higher-order terms that are of interest here are those involving the tensor coupling to the neutral
vector meson; these terms incorporate the tensor force into the calculation, which is known to play
a crucial role in these systems \cite{ref:Mi85,ref:Mi05} and does not enter in leading-order 
contributions \cite{ref:Mc04}. 

In recent years, a number of new avenues have opened to study hypernuclei with increased accuracy. Of
particular interest to this work are $\gamma$-ray coincidence and $(e,e'K^{+})$ experiments.
Recent $\gamma$-ray experiments have measured the fine structure of a number of light hypernuclei \cite{ref:Ta05,ref:Uk05},
including the measurement of the ground-state particle-hole splitting in $_{\Lambda}^{16}$O.
High precision $(e,e'K^{+})$ experiments have measured, or are set to measure, a number of similar states
in light hypernuclei \cite{ref:Ur01,ref:Mi03}, including the ground-states of $_{\Lambda}^{12}$B and 
$_{\Lambda}^{16}$N. Unfortunately, most of the states that have been measured with these techniques thus far 
lie below the range of $A$ accessible to the Kohn-Sham approach used in this work; therefore, the experimental 
constraints on heavier hypernuclei are confined to upper bounds provided by $\left(\pi^{+},K^{+}\right)$ reactions 
\cite{ref:Pi91,ref:Be79,ref:Ha96}. It is this region of medium to heavy single $\Lambda$-hypernuclei that this 
work seeks to investigate.

A number of recent calculations have tackled this problem. Shell model calculations in p-shell hypernuclei 
were conducted using two-body matrix elements, accurately describing the known data \cite{ref:Mi85,ref:Mi05}. 
The influence of zero-range effective $\Lambda$NN interactions on p-shell hypernuclei has also been investigated \cite{ref:Fe01}.
Another model of interest uses strangeness changing response functions to calculate the spectra of $^{16}_{Y}$O and
$^{40}_{Y}$Ca; the resulting ground-state particle-hole splittings are small \cite{ref:Mu01}. The spectra
of $_{\Lambda}^{16}$O has also been calculated from a folded diagram method using realistic hyperon-nucleon 
potentials \cite{ref:Tz99}.

In Section \ref{sec:doub}, we develop a method to calculate the particle-hole matrix elements of interest here. We then
present the results of this analysis and make some conclusions in Section \ref{sec:result}.

\section{$\mathbf{s_{1/2}}$-doublets}
\label{sec:doub}

Consider nuclei like $_{\Lambda}^{16}$O; the ground-states of such systems are, in fact, 
particle-hole states. One process by which nuclei of this type are created is the reaction 
$\left(\pi^{+},K^{+}\right)$ on target nuclei with closed proton and neutron shells
\cite{ref:Pi91,ref:Be79,ref:Ha96}. During the course of this reaction a neutron is converted 
into a $\Lambda$. As a result, a neutron hole is also created which, for the ground-state, inhabits 
the outermost neutron shell. The angular momentum of the $\Lambda$ and the neutron hole
couple to form a multiplet. However, due to the fact that in the ground-state the $\Lambda$ occupies 
the $1\mathrm{s}_{1/2}$ shell, there are only two states in these multiplets. It is these 
configurations that we refer to as s$_{1/2}$-doublets. The reaction 
$(e,e'K^{+})$ is another process used to create nuclei of this type 
\cite{ref:Ur01,ref:Mi03}. This process differs in that a proton hole is created here and that 
greater resolution is possible.

In order to calculate the size of these splittings, we must first construct an effective interaction to model
these systems. The procedure we follow here is similar to the method developed by Machleidt and others \cite{ref:Ma86}. 
In this scheme, the effective NN interaction is represented by the exchange of mesons; then, a hierarchy 
of Feynman diagrams depicting all the possible interactions is developed to reproduce the NN interaction. The contribution for a given 
diagram is just the product of the vertex contributions and the meson propagator, or
\begin{equation}
V(q_{\mu})=\left(\overline{U}_{\alpha}\Gamma_{1}U_{\alpha}\right)D(q_{\mu})
\left(\overline{U}_{\alpha'}\Gamma_{2}U_{\alpha'}\right)
\label{eqn:INT1}
\end{equation}

\noindent where $U_{\alpha}$ are the Dirac free fields, $D(q_{\mu})$ is the meson propagator, and the 
vertex factors, $\Gamma_{1}$ and $\Gamma_{2}$, are taken directly from the underlying effective lagrangian 
through the relation
\begin{equation}
\delta L_{i}=i\overline{\psi}\Gamma_{i}\psi\phi_{\beta}
\end{equation}

\noindent where $\phi_{\beta}$ is some meson field operator. However, nuclei are comprised of bound nucleons
and not free fields. Therefore, in the present calculation, we improve on this system by replacing the free fields 
with the Kohn-Sham (or Hartree) wave functions \cite{ref:Fu97}
\begin{equation}
\psi_{\alpha}=\psi_{n\kappa mt}=\frac{1}{r}\left(\begin{array}{c} iG_{n\kappa m}(r)\Phi_{\kappa m}(\theta,\phi) \\ 
-F_{n\kappa m}(r)\Phi_{-\kappa m}(\theta,\phi) \end{array} \right)\zeta_{t}
\label{eqn:HAR1}
\end{equation}

\noindent The contribution for any given diagram now takes the form (following from Eq.\ (\ref{eqn:INT1}))
\begin{equation}
V(q_{\mu})=\left(\overline{\psi}_{\alpha}\Gamma_{1}\psi_{\alpha}\right)D(q_{\mu})
\left(\overline{\psi}_{\alpha'}\Gamma_{2}\psi_{\alpha'}\right)
\label{eqn:INT2}
\end{equation}

\noindent where we now define the effective interaction in momentum space as 
\begin{equation}
\overline{V}_{EFT}(q_{\mu})\equiv\Gamma_{1}D(q_{\mu})\Gamma_{2}
\label{eqn:HAR2}
\end{equation}

\noindent and $V_{EFT}(q_{\mu})= \gamma_{4}^{(1)}\gamma_{4}^{(2)}\overline{V}_{EFT}(q_{\mu})$.
The meson propagators relevant to this analysis are (here the conventions of \cite{ref:Wa05} 
are used)
\begin{eqnarray}
\frac{1}{i}\frac{1}{q_{\mu}^{2}+m_{S}^{2}} \;\;\; & ; & \;\;\; scalar \nonumber \\ 
\frac{1}{i}\frac{1}{q_{\mu}^{2}+m_{V}^{2}}
\left(\delta_{\mu\nu}+\frac{q_{\mu}q_{\nu}}{m_{V}^{2}}\right) \;\;\; & ; & \;\;\; vector
\end{eqnarray}

\noindent The second term in the vector meson propagator vanishes in any calculation due to conservation
of the baryon current (as a result, it is henceforth neglected). The relevant nucleon vertex factors are
\begin{eqnarray}
ig_{S} \;\;\; & ; & \;\;\; scalar \nonumber \\ 
-g_{V}\gamma_{\mu} \;\;\; & ; & \;\;\; vector \nonumber \\
-\frac{f_{V}g_{V}}{4M}\sigma_{\mu\nu}q_{\mu} \;\;\; & ; & \;\;\; vector \; tensor
\end{eqnarray}

\noindent (Note that the $\Lambda$ vertex factors are equivalent to their nucleon counterparts except 
for the coupling constants \cite{ref:Mc04}.) Now a series of diagrams can be written down to represent 
the NN interaction, each of which will contribute to this effective interaction in the form of Eq.\ 
(\ref{eqn:HAR2}). Once a Fourier transform is performed on the effective interaction, it can be used to 
calculate the two-body matrix elements that govern the particle-hole interactions of interest here.

Now let us return to the case of single $\Lambda$-hypernuclei. The effective $\Lambda$N interaction is determined
via the method outlined above and follows directly from our underlying effective lagrangian 
(see \cite{ref:Mc04}). In this case, no single isovector meson coupling is allowed;
as a result, we confine the following discussion to isoscalar, scalar and vector, exchange.
As a first approximation, we include only the leading-order contributions arising from contact vertices; 
this corresponds to simple scalar and neutral vector couplings. To acquire each portion of the effective 
interaction, we simply take the product of the nucleon vertex factor, 
the meson propagator, and the $\Lambda$ vertex factor.
The simple scalar and neutral vector exchange contributions to the effective interaction take the form
\begin{equation}
\overline{V}_{S}(q_{\mu})=\left(ig_{S}\right)\left(\frac{1}{i}\frac{1}{q_{\mu}^{2}+m_{S}^{2}}\right)
\left(ig_{S\Lambda}\right)
\label{eqn:SC1}
\end{equation}
\begin{equation}
\overline{V}_{V}(q_{\mu})=\left(-g_{V}\gamma_{\mu}^{(1)}\right)\left(\frac{1}{i}\frac{1}{q_{\mu}^{2}+m_{V}^{2}}\right)
\left(-g_{V\Lambda}\gamma_{\mu}^{(2)}\right)
\label{eqn:VE1}
\end{equation}

\noindent Next we take the Fourier transform of Eqs.\ (\ref{eqn:SC1}) and (\ref{eqn:VE1}), neglecting retardation in the 
meson propagator (i.e.\ $q_{4} \rightarrow 0$), and we get
\begin{equation}
\overline{V}_{S}(r_{12})=-\frac{g_{S}g_{S\Lambda}}{4\pi}\frac{e^{-m_{S}r_{12}}}{r_{12}}
\end{equation}
\begin{eqnarray}
\overline{V}_{V}(r_{12}) & = & \frac{g_{V}g_{V\Lambda}}{4\pi}\gamma_{\mu}^{(1)}\gamma_{\mu}^{(2)}
\frac{e^{-m_{V}r_{12}}}{r_{12}} \nonumber \\ & = & \overline{V}_{V(1)}(r_{12})\gamma_{4}^{(1)}\gamma_{4}^{(2)}
+\overline{V}_{V(2)}(r_{12})\gamma_{i}^{(1)}\gamma_{i}^{(2)}
\end{eqnarray}

\noindent where $r_{12}=(r_{1}^{2}+r_{2}^{2}-2r_{1}r_{2}\cos\theta_{12})^{1/2}$. Notice that this
corresponds to simple Yukawa couplings to both the scalar and neutral vector mesons. The effective interaction
to this order was used in calculations conducted in \cite{ref:Mc04}; unfortunately, it proved inadequate to 
fully describe the ground-state splittings in single $\Lambda$-hypernuclei. As it turns out, the effective 
interaction to this order includes only the spin-spin force. However, this neglects the fact that the tensor 
force is known to play a crucial role in these systems \cite{ref:Mi85,ref:Mi05}. Therefore, the natural extension of 
this approach is to include the higher-order contributions involving tensor couplings and thereby incorporate the 
tensor force into the calculation. 

There are three higher-order terms containing gradient vertices (or tensor couplings) of the neutral vector meson; their resulting 
contributions to the effective interaction are now considered. The contributions from neutral vector meson exchange with a tensor 
coupling on one vertex to the effective interaction are
\begin{eqnarray}
\overline{V}_{NT}(q_{\mu}) & = & \left(-\frac{f_{V}g_{V}}{4M}\sigma_{\mu\nu}^{(1)}q_{\mu}\right)
\left(\frac{1}{i}\frac{1}{q_{\nu}^{2}+m_{V}^{2}}\right)\left(-g_{V\Lambda}\gamma_{\nu}^{(2)}\right)
\nonumber \\ & & +\left(\frac{f_{V}g_{V}}{4M}\sigma_{\mu\nu}^{(1)}q_{\nu}\right)
\left(\frac{1}{i}\frac{1}{q_{\mu}^{2}+m_{V}^{2}}\right)\left(-g_{V\Lambda}\gamma_{\mu}^{(2)}\right)
\label{eqn:NT1}
\end{eqnarray}
\begin{eqnarray}
\overline{V}_{\Lambda T}(q_{\mu}) & = & \left(-g_{V}\gamma_{\mu}^{(1)}\right)
\left(\frac{1}{i}\frac{1}{q_{\mu}^{2}+m_{V}^{2}}\right)
\left(-\frac{g_{T\Lambda}g_{V}}{4M}\sigma_{\mu\nu}^{(2)}q_{\nu}\right)
\nonumber \\ & & +\left(-g_{V}\gamma_{\nu}^{(1)}\right)
\left(\frac{1}{i}\frac{1}{q_{\nu}^{2}+m_{V}^{2}}\right)
\left(\frac{g_{T\Lambda}g_{V}}{4M}\sigma_{\mu\nu}^{(2)}q_{\mu}\right)
\label{eqn:LT1}
\end{eqnarray}

\noindent where the contribution with the tensor coupling on the nucleon ($\Lambda$) vertex is denoted by $\overline{V}_{NT}$ 
($\overline{V}_{\Lambda T}$). Notice that two terms arise in both $\overline{V}_{NT}$ and
$\overline{V}_{\Lambda T}$ due to the fact that the quantity $V_{\mu\nu}=\partial_{\mu}V_{\nu}-\partial_{\nu}V_{\mu}$ 
appears in the lagrangian \cite{ref:Mc04}. Now the Fourier transforms of Eqs.\ 
(\ref{eqn:NT1}) and (\ref{eqn:LT1}), neglecting retardation in the meson propagator, yield respectively
\begin{eqnarray}
\overline{V}_{NT}(r_{12}) & = & \frac{if_{V}g_{V}g_{V\Lambda}}{4\pi}\frac{m_{V}}{2M}
\left\{i\left[\left(\sigma^{(1)} \times \hat{r}\right) \cdot \sigma^{(2)}\right]\gamma_{4}^{(2)}\gamma_{5}^{(2)}\right. 
\nonumber \\ & & \left. -\left(\sigma^{(1)} \cdot \hat{r}\right)\gamma_{5}^{(1)}\gamma_{4}^{(2)}\right\}
\left(1+\frac{1}{m_{V}r_{12}}\right)\frac{e^{-m_{V}r_{12}}}{r_{12}} \nonumber \\ & = &
\overline{V}_{NT(1)}(r_{12})\gamma_{4}^{(2)}\gamma_{5}^{(2)}
+\overline{V}_{NT(2)}(r_{12})\gamma_{5}^{(1)}\gamma_{4}^{(2)}
\end{eqnarray}
\begin{eqnarray}
\overline{V}_{\Lambda T}(r_{12}) & = & \frac{ig_{T\Lambda}g_{V}^{2}}{4\pi}\frac{m_{V}}{2M}
\left\{i\left[\sigma^{(1)} \cdot \left(\sigma^{(2)} \times \hat{r}\right)\right]\gamma_{4}^{(1)}\gamma_{5}^{(1)}\right. 
\nonumber \\ & & \left. -\left(\sigma^{(2)} \cdot \hat{r}\right)\gamma_{4}^{(1)}\gamma_{5}^{(2)}\right\}
\left(1+\frac{1}{m_{V}r_{12}}\right)\frac{e^{-m_{V}r_{12}}}{r_{12}}\nonumber \\ & = &
\overline{V}_{\Lambda T(1)}(r_{12})\gamma_{4}^{(1)}\gamma_{5}^{(1)}
+\overline{V}_{\Lambda T(2)}(r_{12})\gamma_{4}^{(1)}\gamma_{5}^{(2)}
\end{eqnarray}

\noindent where the following relation
\begin{eqnarray}
\sigma_{\mu\nu}^{(1)}q_{\mu} & = & \sigma_{ij}^{(1)}q_{i} + \sigma_{i4}^{(1)}q_{i} 
+ \sigma_{4j}^{(1)}q_{4} \nonumber \\ & = & \vec{\sigma}^{(1)} \times \vec{q}
- \vec{\sigma}^{(1)} \cdot \vec{q} \; \gamma_{5}^{(1)}
\end{eqnarray}

\noindent has been used (and also for $(1) \leftrightarrow (2)$). It is interesting to note that in \cite{ref:Ma86},
the terms corresponding to $\overline{V}_{NT(1)}$ and $\overline{V}_{\Lambda T(1)}$ both develop into a 
combination of spin-spin and tensor forces while the terms corresponding to $\overline{V}_{NT(2)}$ and 
$\overline{V}_{\Lambda T(2)}$ both form a combination of central and spin-orbit forces. 

For vector meson exchange with a tensor coupling on both vertices, the effective interaction is
\begin{eqnarray}
\overline{V}_{TT}(q_{\mu}) & = & \left(-\frac{f_{V}g_{V}}{4M}\sigma_{\mu\nu}^{(1)}q_{\mu}\right)
\left(\frac{1}{i}\frac{1}{q_{\nu}^{2}+m_{V}^{2}}\right)
\left(\frac{g_{T\Lambda}g_{V}}{4M}\sigma_{\mu\nu}^{(2)}q_{\mu}\right)
\nonumber \\ & & +\left(\frac{f_{V}g_{V}}{4M}\sigma_{\mu\nu}^{(1)}q_{\nu}\right)
\left(\frac{1}{i}\frac{1}{q_{\mu}^{2}+m_{V}^{2}}\right)
\left(-\frac{g_{T\Lambda}g_{V}}{4M}\sigma_{\mu\nu}^{(2)}q_{\nu}\right)
\label{eqn:TT1}
\end{eqnarray}

\noindent Taking the Fourier transform of Eq.\ (\ref{eqn:TT1}), again neglecting retardation in the 
meson propagator, gives
\begin{eqnarray}
\overline{V}_{TT}(r_{12}) & = & \frac{f_{V}g_{T\Lambda}g_{V}^{2}}{12\pi}\frac{m_{V}^{2}}{8M^{2}}
\left[\sigma^{(1)} \cdot \sigma^{(2)} \left(2 + \gamma_{5}^{(1)}\gamma_{5}^{(2)}\right)\right.
\nonumber \\ & & -\left. S_{12}(\hat{r}_{12})\left(1 - \gamma_{5}^{(1)}\gamma_{5}^{(2)}\right)
\left(1+\frac{3}{m_{V}r_{12}}+\frac{3}{m_{V}^{2}r_{12}^{2}}\right)\right]
\frac{e^{-m_{V}r_{12}}}{r_{12}} \nonumber \\ & = &
\overline{V}_{TT(1)}(r_{12})\left(2 + \gamma_{5}^{(1)}\gamma_{5}^{(2)}\right) 
+\overline{V}_{TT(2)}(r_{12})\left(1 - \gamma_{5}^{(1)}\gamma_{5}^{(2)}\right)
\end{eqnarray}

\noindent where
\begin{equation}
S_{12}(\hat{r})=3(\vec{\sigma}^{(1)} \cdot \hat{r})(\vec{\sigma}^{(2)} \cdot \hat{r}) - 
\vec{\sigma}^{(1)} \cdot \vec{\sigma}^{(2)}
\end{equation}

Lastly, we combine the interactions from all five contributions into a single effective interaction, or
\begin{eqnarray}
V(r_{12}) & = & \gamma_{4}^{(1)}\gamma_{4}^{(2)}\overline{V}(r_{12}) \nonumber \\ & = & 
\gamma_{4}^{(1)}\gamma_{4}^{(2)}\left(\overline{V}_{S}+\overline{V}_{V}+\overline{V}_{NT}+\overline{V}_{\Lambda T}
+\overline{V}_{TT}\right) \nonumber \\ & = & \overline{V}_{S}\gamma_{4}^{(1)}\gamma_{4}^{(2)}+\overline{V}_{V(1)}
+\overline{V}_{V(2)}\gamma_{4}^{(1)}\gamma_{i}^{(1)}\gamma_{4}^{(2)}\gamma_{i}^{(2)} 
+\overline{V}_{NT(1)}\gamma_{4}^{(2)}\gamma_{5}^{(2)} \nonumber \\ & &
+ \overline{V}_{NT(2)}\gamma_{5}^{(1)}\gamma_{4}^{(2)}
+\overline{V}_{\Lambda T(1)}\gamma_{4}^{(1)}\gamma_{5}^{(1)} 
+\overline{V}_{\Lambda T(2)}\gamma_{4}^{(1)}\gamma_{5}^{(2)}\nonumber \\ & &
+\overline{V}_{TT(1)}\left(2\gamma_{4}^{(1)}\gamma_{4}^{(2)} 
+ \gamma_{4}^{(1)}\gamma_{5}^{(1)}\gamma_{4}^{(2)}\gamma_{5}^{(2)}\right) \nonumber \\ & &
+\overline{V}_{TT(2)}\left(\gamma_{4}^{(1)}\gamma_{4}^{(2)} 
- \gamma_{4}^{(1)}\gamma_{5}^{(1)}\gamma_{4}^{(2)}\gamma_{5}^{(2)}\right)
\label{eqn:totint}
\end{eqnarray}

Now that we have constructed an effective interaction, it can be used to determine the particle-hole splittings.
In order to accomplish this, we must first calculate matrix elements
of the following varieties
\begin{equation}
\langle (n_{1}l_{1}j_{1})(n_{2}l_{2}j_{2})JM|
V_{i}(r_{12})|(n_{3}l_{3}j_{3})(n_{4}l_{4}j_{4})J'M'\rangle
\label{eqn:A1}
\end{equation}
\begin{equation}
\langle (n_{1}l_{1}j_{1})(n_{2}l_{2}j_{2})JM|
V_{i}(r_{12})\vec{\sigma}^{(1)}\cdot\vec{\sigma}^{(2)}
|(n_{3}l_{3}j_{3})(n_{4}l_{4}j_{4})J'M'\rangle
\label{eqn:A2}
\end{equation}
\begin{equation}
\langle (n_{1}l_{1}j_{1})(n_{2}l_{2}j_{2})JM|
V_{i}(r_{12})(\sigma^{(1)} \cdot \hat{r})|(n_{3}l_{3}j_{3})(n_{4}l_{4}j_{4})J'M'\rangle
\label{eqn:A3}
\end{equation}
\begin{equation}
\langle (n_{1}l_{1}j_{1})(n_{2}l_{2}j_{2})JM|
V_{i}(r_{12})i\left[\left(\sigma^{(1)} \times \hat{r}\right) \cdot \sigma^{(2)}\right]
|(n_{3}l_{3}j_{3})(n_{4}l_{4}j_{4})J'M'\rangle
\label{eqn:A4}
\end{equation}

\noindent where the single-particle wave functions are specified by $\{nlj\}$, corresponding to either the upper or 
lower components in Eq.\ (\ref{eqn:HAR1}), and $V_{i}(r_{12})$ is some part of the effective interaction.
Next, we expand each part of this effective interaction in terms of Legendre polynomials \cite{ref:Fu85}
\begin{eqnarray}
V_{i}(r_{12}) & = & \sum_{k=0}^{\infty}f_{k}^{i}(r_{1},r_{2}) P_{k}(\cos\theta_{12}) \label{eqn:A8}
\\ & = & \sum_{k=0}^{\infty}f_{k}^{i}(r_{1},r_{2})C_{k}(1)\cdot C_{k}(2)
\end{eqnarray}

\noindent where 
\begin{equation}
C_{kq} = \left(\frac{4\pi}{2k+1}\right)^{1/2}Y_{kq}(\theta,\phi) 
\end{equation}

\noindent Inverting Eq.\ (\ref{eqn:A8}) yields the expression
\begin{equation}
f_{k}^{i}(r_{1},r_{2}) = \frac{2k+1}{2}\int_{-1}^{1}d(\cos\theta_{12})P_{k}(\cos\theta_{12})V_{i}(r_{12}) 
\label{eqn:A9} 
\end{equation}

\noindent In the case of Eq.\ (\ref{eqn:A2}), the effective interaction is coupled to Pauli matrices. 
Therefore, Eq.\ (\ref{eqn:A8}) is modified to
\begin{equation}
V_{i}(r_{12})\vec{\sigma}^{(1)}\cdot\vec{\sigma}^{(2)}=\sum_{k\lambda}(-1)^{k+1-\lambda}f_{k}^{i}(r_{1},r_{2})
\chi_{\lambda}^{(k,1)}(1)\cdot\chi_{\lambda}^{(k,1)}(2)
\end{equation}

\noindent Here $\chi_{\lambda\mu}^{(k,1)}$ are $C_{kq}$ coupled to Pauli matrices, shown by
\begin{equation}
\chi_{\lambda\mu}^{(k,1)}=\sum_{qq'}\mathrm{C}_{kq}\sigma_{1q'}\langle kq1q'|k1\lambda\mu\rangle
\end{equation}

Next, we reduce Eq.\ (\ref{eqn:A3}) to the form of Eq.\ (\ref{eqn:A1}) (up to a sign). This is possible 
as the operator $(\sigma^{(1)} \cdot \hat{r})$ acts only on the angular portion of the Hartree wave 
functions. Using the relation
\begin{equation}
(\sigma^{(1)} \cdot \hat{r})\Phi_{\kappa_{1}m_{1}}^{(1)} = -\Phi_{-\kappa_{1}m_{1}}^{(1)}
\end{equation}

\noindent the expression Eq.\ (\ref{eqn:A3}) can be rewritten in the following form
\begin{eqnarray}
& \langle (n_{1}l_{1}j_{1})(n_{2}l_{2}j_{2})JM|V_{i}(r_{12})(\sigma^{(1)} \cdot \hat{r})
|(n_{3}l_{3}j_{3})(n_{4}l_{4}j_{4})J'M'\rangle & \nonumber \\ = & - \langle (n_{1}l_{1}j_{1})(n_{2}l_{2}j_{2})JM|
V_{i}(r_{12})|(n_{3}[l_{3A} \leftrightarrow l_{3B}]j_{3})(n_{4}l_{4}j_{4})J'M'\rangle & 
\label{eqn:ME1}
\end{eqnarray}

\noindent where $l_{iA}$ and $l_{iB}$ are the $l$ values corresponding to the upper and lower 
Hartree spinors respectively for the $i$th wave function. Eq.\ (\ref{eqn:ME1}) is readily
generalized to the case $(1) \leftrightarrow (2)$.

Similarly, Eq.\ (\ref{eqn:A4}) can be reduced to the form of Eq.\ (\ref{eqn:A1}) (up to a factor). 
Here we employ the relation
\begin{eqnarray}
\lefteqn{i\left[\left(\sigma^{(1)} \times \hat{r}\right) \cdot \sigma^{(2)}\right]
|(l_{1}\frac{1}{2}j_{1})(l_{2}\frac{1}{2}j_{2})JM\rangle} \nonumber \\ & = & 
\sqrt{2}\sum_{l_{1}^{'}}\sum_{j_{1}^{'}j_{2}^{'}}
|(l_{1}^{'}\frac{1}{2}j_{1}^{'})(l_{2}\frac{1}{2}j_{2}^{'})JM\rangle
(-1)^{j_{1}+j_{2}^{'}+J}
\left\{\begin{array}{ccc} J & j_{2}^{'} & j_{1}^{'} \\ 1 & j_{1} & j_{2} \\ \end{array} \right\} \nonumber \\ & & \times 
\langle l_{1}^{'}\frac{1}{2}j_{1}^{'}||\left(\vec{\sigma}^{(1)}\times \hat{r}\right)||l_{1}\frac{1}{2}j_{1}\rangle 
\langle l_{2}\frac{1}{2}j_{2}^{'}||\vec{\sigma}^{(2)}||(l_{2}\frac{1}{2})j_{2}\rangle
\end{eqnarray} 

\noindent We use \cite{ref:Ed57} to further simplify the reduced matrix elements. Now we can write
all possible matrix elements to this order in terms of Eqs.\ (\ref{eqn:A1}) and (\ref{eqn:A2}).

The matrix elements in Eqs.\ (\ref{eqn:A1}) and (\ref{eqn:A2}) are actually six dimensional integrals.  
Treating the $\gamma$-matrices as $2 \times 2$ block matrices operating on the upper and lower components of the Hartree
spinors, these Dirac matrix elements, for each term in the interaction, are actually the sum of
four separate integrals. Thankfully, angular momentum relations allow one to integrate out the angular dependence 
\cite{ref:Ed57}. Eq.\ (\ref{eqn:A1}) becomes
\begin{eqnarray}
(21) & = & {\displaystyle\sum_{k=0}^{\infty}} \langle12|f_{k}^{i}(r_{1},r_{2})|34\rangle
(-1)^{j_{2}+j_{3}+J} \left\{ \begin{array}{ccc} J & j_{2} & j_{1} \\ k & j_{3} & j_{4} \\ \end{array} \right\} 
\delta_{JJ'}\delta_{MM'} \nonumber \\ & & \times 
\langle (l_{1}\frac{1}{2})j_{1}||C_{k}(1)||(l_{3}\frac{1}{2})j_{3}\rangle
\langle(l_{2}\frac{1}{2})j_{2}||C_{k}(2)||(l_{4}\frac{1}{2})j_{4}\rangle
\label{eqn:A10}
\end{eqnarray}

\noindent and Eq.\ (\ref{eqn:A2}) becomes
\begin{eqnarray}
(22) & = & {\displaystyle\sum_{k=0}^{\infty}\sum_{\lambda}}\langle12|f_{k}^{i}(r_{1},r_{2})
|34\rangle(-1)^{j_{2}+j_{3}+J+k+1-\lambda} \left\{ \begin{array}{ccc} J & j_{2} & j_{1} \\ 
\lambda & j_{3} & j_{4} \\ \end{array} \right\} \delta_{JJ'}\delta_{MM'} \nonumber \\ & & \times 
\langle(l_{1}\frac{1}{2})j_{1}||\chi_{\lambda}^{(k,1)}(1)||(l_{3}\frac{1}{2})j_{3}\rangle
\langle(l_{2}\frac{1}{2})j_{2}||\chi_{\lambda}^{(k,1)}(2)||(l_{4}\frac{1}{2})j_{4}\rangle
\label{eqn:A11}
\end{eqnarray}

\noindent where $i$ denotes some portion of the effective interaction. The 6-$j$ symbols limit the possible 
allowed values of $k$ and $\lambda$. The reduced matrix elements are evaluated using \cite{ref:Ed57} 
and further limit $k$ and $\lambda$. 

Now consider the remaining two-dimensional radial integrals, where the numbers are a shorthand for all 
the quantum numbers needed to uniquely specify the radial wave function \cite{ref:Fu85},
\begin{equation}
\langle12|f_{k}^{i}(r_{1},r_{2})|34\rangle = \int_{0}^{\infty}\!\!\int_{0}^{\infty} 
dr_{1}dr_{2}U_{1}(r_{1})U_{2}(r_{2})f_{k}^{i}(r_{1},r_{2})U_{3}(r_{1})U_{4}(r_{2})
\end{equation}

\noindent Here $R(r) = U(r) / r$ are the appropriate radial Dirac wave functions, in terms of $G_{a}(r)$ 
and $F_{a}(r)$. Note that as the upper and lower Hartree spinors have different $l$ values, the reduced 
matrix elements in Eqs.\ (\ref{eqn:A10}) and (\ref{eqn:A11}) must have the corresponding, appropriate $l$ values.

Using the Hartree spinor representation, the particle-hole matrix element is expressed as 
a sum of Dirac matrix elements of the types shown above \cite{ref:Fe71}
\begin{equation}
v_{ab;lm}^{J}=\sum_{J'}(2J'+1) \left\{ \begin{array}{ccc}
j_{m} & j_{a} & J' \\ j_{b} & j_{l} & J \\ \end{array} \right\}
\langle lbJ'|\gamma_{4}^{(1)}\gamma_{4}^{(2)}\overline{V}|amJ'\rangle
\end{equation} 

\noindent No exchange term is required here, since the $\Lambda$ and the nucleon are distinguishable 
particles. For example, the particle-hole matrix element for the $\overline{V}_{V(2)}$ is
\begin{eqnarray}
\lefteqn{v_{32;14}^{J}(\overline{V}_{V(2)})
= (-1)^{j_{2}+j_{3}+J}\sum_{k}^{\infty}\sum_{\lambda}(-1)^{k}} \nonumber \\ & & 
\times \left\{\begin{array}{ccc} j_{2} & j_{4} & \lambda \\ j_{1} & j_{3} & J \\ \end{array} \right\}
\int\int dr_{1}dr_{2} \left\{ G_{1}(r_{1})F_{3}(r_{1})f_{k}^{V(2)}(r_{1},r_{2})G_{2}(r_{2})F_{4}(r_{2}) \right. 
\nonumber \\ & & \times \langle(l_{1A}\frac{1}{2})j_{1}||\chi_{\lambda}^{(k,1)}(1)||(l_{3B}\frac{1}{2})j_{3}\rangle
\langle(l_{2A}\frac{1}{2})j_{2}||\chi_{\lambda}^{(k,1)}(2)||(l_{4B}\frac{1}{2})j_{4}\rangle \nonumber \\ & & 
-G_{1}(r_{1})F_{3}(r_{1})f_{k}^{V(2)}(r_{1},r_{2})F_{2}(r_{2})G_{4}(r_{2}) \nonumber \\ & & \times 
\langle(l_{1A}\frac{1}{2})j_{1}||\chi_{\lambda}^{(k,1)}(1)||(l_{3B}\frac{1}{2})j_{3}\rangle
\langle(l_{2B}\frac{1}{2})j_{2}||\chi_{\lambda}^{(k,1)}(2)||(l_{4A}\frac{1}{2})j_{4}\rangle \nonumber \\ & &
-F_{1}(r_{1})G_{3}(r_{1})f_{k}^{V(2)}(r_{1},r_{2})G_{2}(r_{2})F_{4}(r_{2}) \nonumber \\ & & \times 
\langle(l_{1B}\frac{1}{2})j_{1}||\chi_{\lambda}^{(k,1)}(1)||(l_{3A}\frac{1}{2})j_{3}\rangle
\langle(l_{2A}\frac{1}{2})j_{2}||\chi_{\lambda}^{(k,1)}(2)||(l_{4B}\frac{1}{2})j_{4}\rangle \nonumber \\ & &
+F_{1}(r_{1})G_{3}(r_{1})f_{k}^{V(2)}(r_{1},r_{2})F_{2}(r_{2})G_{4}(r_{2}) \nonumber \\ & & \left. \times 
\langle(l_{1B}\frac{1}{2})j_{1}||\chi_{\lambda}^{(k,1)}(1)||(l_{3A}\frac{1}{2})j_{3}\rangle
\langle(l_{2B}\frac{1}{2})j_{2}||\chi_{\lambda}^{(k,1)}(2)||(l_{4A}\frac{1}{2})j_{4}\rangle \right\}
\label{eqn:ph1}
\end{eqnarray} 

\noindent Now the splitting, for a $\mathrm{s}_{1/2}$-doublet, is just the difference between the particle-hole matrix 
elements of the two available states, or
\begin{equation}
\delta\epsilon=v_{n\Lambda;n\Lambda}^{J=j_{1}+j_{2}}-v_{n\Lambda;n\Lambda}^{J=|j_{1}-j_{2}|}
\label{eqn:spl}
\end{equation}

\noindent The substitutions used to get the appropriate indices for this case are $n=1,3$ and $\Lambda = 2,4$.
The solution to the Kohn-Sham equations yields a single-particle energy level for the ground-state, 
$E_{\Lambda}$. As previously mentioned, this level is in fact a doublet; however, Eq.\ (\ref{eqn:spl}) 
determines only the size of the splitting. In order to determine the position of the doublet relative to 
$E_{\Lambda}$, one needs the relation
\begin{equation}
\sum_{J}\left(2J+1\right)\delta\epsilon=0
\label{eqn:pos}
\end{equation}

We now have a framework with which to calculate the size of the s$_{1/2}$-doublets
of the single $\Lambda$-hypernuclei of interest here and to determine their location relative to $E_{\Lambda}$. 
The problem is reduced to Slater integrals and some algebra; the 6-$j$ and 9-$j$ symbols are determined using 
\cite{ref:Ma55,ref:Ro59}. The Dirac wave functions needed to solve the integrals are taken as the solutions to 
the radial Kohn-Sham equations \cite{ref:Mc04}. Once all the parameters in the underlying lagrangian are
fixed, the splitting is completely determined in this approach. We also mention that this approach is 
applicable to excited states and multiplets for this class of nuclei.

\section{Results}
\label{sec:result}
\begin{table}
\begin{center}
\begin{tabular}{|c|c|c|c|c|c|c|c|c|} \hline\hline
Nucleus & State & Levels & $V(2)$ & $NT(1)$ & $\Lambda T(1)$ & $TT(1)$ & $TT(2)$ & $\delta\epsilon$ \\ \hline
$^{12}_{\Lambda}$B   & $(1p_{3/2})_{p}^{-1}(1s_{1/2})_{\Lambda}$  & $2^{-}_{GS}$, $1^{-}$ & 
-425 & -74  & -185   & -1068 & -258 & -2011 \\ \hline
$^{16}_{\Lambda}$N   & $(1p_{1/2})_{p}(1s_{1/2})_{\Lambda}$       & $0^{-}_{GS}$, $1^{-}$ & 
-476 & 283  & -1052  & 476   & 791  & 23    \\ \hline
                     & $(1p_{3/2})_{p}^{-1}(1s_{1/2})_{\Lambda}$  & $2^{-}_{LL}$, $1^{-}$ & 
-314 & -57  & -146   & -901  & -212 & -1632 \\ \hline
$^{16}_{\Lambda}$O   & $(1p_{1/2})_{n}(1s_{1/2})_{\Lambda}$       & $0^{-}_{GS}$, $1^{-}$ & 
-484 & 287  & -1071  & 485   & 805  & 22    \\ \hline
$^{28}_{\Lambda}$Si  & $(1d_{5/2})_{n}^{-1}(1s_{1/2})_{\Lambda}$  & $3^{+}_{GS}$, $2^{+}$ & 
-299 & -23  & -49    & -490  & -150 & -1011 \\ \hline
$^{32}_{\Lambda}$S   & $(2s_{1/2})_{n}(1s_{1/2})_{\Lambda}$       & $1^{+}_{GS}$, $0^{+}$ & 
-223 & -174 & -631   & -1034 & -198 & -2260 \\ \hline
$^{40}_{\Lambda}$Ca  & $(1d_{3/2})_{n}^{-1}(1s_{1/2})_{\Lambda}$  & $1^{+}_{GS}$, $2^{+}$ & 
-308 & 34   & -149   & 277   & 252  & 107   \\ \hline
                     & $(1d_{3/2})_{n}^{-1}(1p_{1/2})_{\Lambda}$  & $2^{-}_{LL}$, $1^{-}$ & 
-376 & 31   & 97     & -385  & -80  & -712  \\ \hline
$^{48}_{\Lambda}$K   & $(1d_{3/2})_{p}^{-1}(1s_{1/2})_{\Lambda}$  & $1^{+}_{GS}$, $2^{+}$ & 
-324 & 35   & -150   & 272   & 247  & 80    \\ \hline
$^{48}_{\Lambda}$Ca  & $(1f_{7/2})_{n}^{-1}(1s_{1/2})_{\Lambda}$  & $4^{-}_{GS}$, $3^{-}$ & 
-147 & -6   & -12    & -223  & -146 & -535  \\ \hline
$^{88}_{\Lambda}$Rb  & $(1f_{5/2})_{p}^{-1}(1s_{1/2})_{\Lambda}$  & $2^{-}_{GS}$, $3^{-}$ & 
-178 & 8    & -38    & 187   & 128  & 108   \\ \hline
$^{88}_{\Lambda}$Sr  & $(1g_{9/2})_{n}^{-1}(1s_{1/2})_{\Lambda}$  & $5^{+}_{GS}$, $4^{+}$ & 
-77  & 0    & 0      & -106  & -67  & -251  \\ \hline
$^{208}_{\Lambda}$Pb & $(1i_{13/2})_{n}^{-1}(1s_{1/2})_{\Lambda}$ & $7^{-}_{GS}$, $6^{-}$ & 
-14  & 0    & -1     & -53   & -37  & -104  \\ \hline
\end{tabular}
\caption{Calculation of the s$_{1/2}$-doublets (and some excited state splittings for $^{16}_{\Lambda}$N
and $^{40}_{\Lambda}$Ca) in select single $\Lambda$-hypernuclei (in keV). Here $\delta\epsilon$
is defined by Eq.\ (\ref{eqn:spl}) and the ground-states are marked by GS (similarly LL denotes lower 
level for the excited states).}
\label{tab:split1}
\end{center}
\end{table}

Here, we discuss the results obtained from the calculation of the 
ground-state particle-hole splittings in single $\Lambda$-hypernuclei by the method 
discussed in the previous section. The goal of this calculation is to evaluate 
$\delta\epsilon$ in Eq.\ (\ref{eqn:spl}). To facilitate this, it is convenient to write $\delta\epsilon$ 
as a sum of the contributions from each portion of the effective interaction, or
\begin{eqnarray}
\delta\epsilon & = & \delta\epsilon[S]+\delta\epsilon[V(1)]+\delta\epsilon[V(2)]
+\delta\epsilon[NT(1)]+\delta\epsilon[NT(2)]+\delta\epsilon[\Lambda T(1)]
\nonumber \\ & & +\delta\epsilon[\Lambda T(2)]+\delta\epsilon[TT(1)]+\delta\epsilon[TT(2)]
\end{eqnarray}

\noindent where these contributions are defined in Eq.\ (\ref{eqn:totint}).
As it turns out, the following terms cancel in the splitting
\begin{equation}
\delta\epsilon[S]=\delta\epsilon[V(1)]=\delta\epsilon[NT(2)]=\delta\epsilon[\Lambda T(2)]=0
\label{eqn:CAN1}
\end{equation}

\noindent This is true for any system in which either the $\Lambda$ or the nucleon hole has 
$j=1/2$. Thus, the total splitting for these states is 
\begin{equation}
\delta\epsilon = \delta\epsilon[V(2)]+\delta\epsilon[NT(1)]+\delta\epsilon[\Lambda T(1)]
+\delta\epsilon[TT(1)]+\delta\epsilon[TT(2)]
\label{eqn:TOT1}
\end{equation}

\noindent It is interesting to note that these terms contribute only spin-spin and tensor forces;
no central or spin-orbit forces survive (see Eq.\ (\ref{eqn:CAN1})). The remaining terms 
can be described in the following fashion: $V(2)$ and $TT(1)$ are purely spin-spin interactions,
$TT(2)$ is purely a tensor interaction, and $NT(1)$ and $\Lambda T(1)$ are both an admixture of spin-spin 
and tensor interactions. Note that the higher-order contributions incorporate the tensor force into the 
calculation; the tensor force is known to play an important role in the $\Lambda$N interaction 
\cite{ref:Mi85,ref:Mi05} and does not appear in the leading-order terms. Now we determine 
the particle-hole matrix elements for each portion of the effective interaction. The two-dimensional 
integrals are calculated numerically using the Hartree spinors, $G_{a}(r)$ and $F_{a}(r)$, acquired by 
solving the self-consistent single-particle equations \cite{ref:Mc04}. The position of the states relative to the Kohn-Sham 
level is determined from Eq.\ (\ref{eqn:pos}). All of the coupling constants used in this calculation
are taken from \cite{ref:Mc04} (specifically the sets G2, which originates from \cite{ref:Fu97}, and M2); 
hence, there are no remaining free parameters.
\begin{figure}
\begin{center}
\includegraphics[scale=.40]{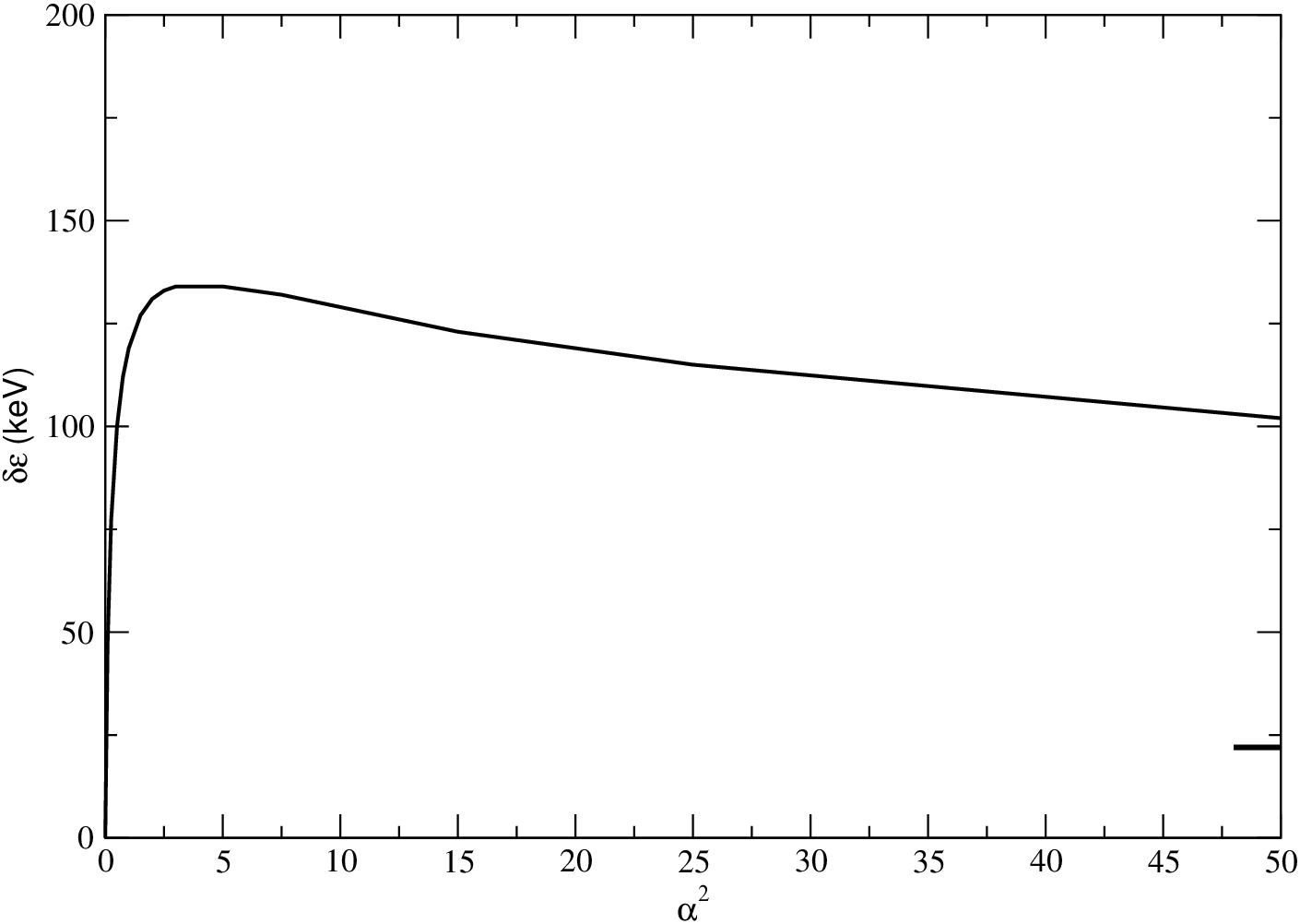}
\caption{Effect of the correlation function from Eq.\ (\ref{eqn:1}) on the total ground-state
splitting for $^{16}_{\Lambda}$O as a function of $\alpha^{2}$. The value of the splitting for
$\alpha^{2}=\infty$ is marked.}
\label{fig:corr1}
\end{center}
\end{figure}
\begin{table}
\begin{center}
\begin{tabular}{|c|c|c|} \hline\hline
Nucleus & $|\delta\epsilon|$ & Reference \\ \hline
$^{12}_{\Lambda}$B   & $\leq$ 140  & \cite{ref:Ju73} \\ \hline
$^{16}_{\Lambda}$O   & +26         & \cite{ref:Ta05,ref:Uk05} \\ \hline
$^{28}_{\Lambda}$Si  & $\leq$ 400  & \cite{ref:Ha96} \\ \hline
$^{32}_{\Lambda}$S   & $\leq$ 1000 & \cite{ref:Be79} \\ \hline
$^{40}_{\Lambda}$Ca  & $\leq$ 2200 & \cite{ref:Pi91} \\ \hline
$^{208}_{\Lambda}$Pb & $\leq$ 2000 & \cite{ref:Ha96} \\ \hline
\end{tabular}
\caption{Experimental constraints on the splittings in keV.}
\label{tab:data1}
\end{center}
\end{table}

The contributions from the surviving portions of the effective interaction, the total splitting, and 
the resulting level orderings for a number of ground-state particle-hole splittings (as well as some excited states) 
are shown in Table (\ref{tab:split1}). Note that the contribution labeled $V(2)$ was the portion of the 
interaction that was investigated in \cite{ref:Mc04}; the interaction to this order failed to reproduce either
the correct level ordering or splitting for the ground-state of $_{\Lambda}^{16}$O. Therefore, the interaction 
was expanded to include the higher-order gradient couplings described above. This expanded interaction, as given by Eq.\ 
(\ref{eqn:totint}), now gives both the proper level ordering and splitting for the ground-state of $_{\Lambda}^{16}$O, 
as shown in Table (\ref{tab:split1}). Note that the inclusion of the tensor force was crucial to achieve the cancellation 
necessary to describe the small experimental splitting in the ground-state of $_{\Lambda}^{16}$O \cite{ref:Ta05,ref:Uk05}, 
in agreement with previous work \cite{ref:Mi05}. 
Similar cancellation occurs for states with $j_{1}+j_{2}+\pi=even$ (where $\pi$ denotes the parity of the 
system). Unfortunately, the splittings shown in Table (\ref{tab:split1}) with $j_{1}+j_{2}+\pi=odd$, the ground-state of 
$_{\Lambda}^{32}$S for instance, are quite large, well outside the known experimental error bars; the experimental 
constraints are listed in Table (\ref{tab:data1}). 
\begin{figure}
\begin{center}
\includegraphics[scale=.40]{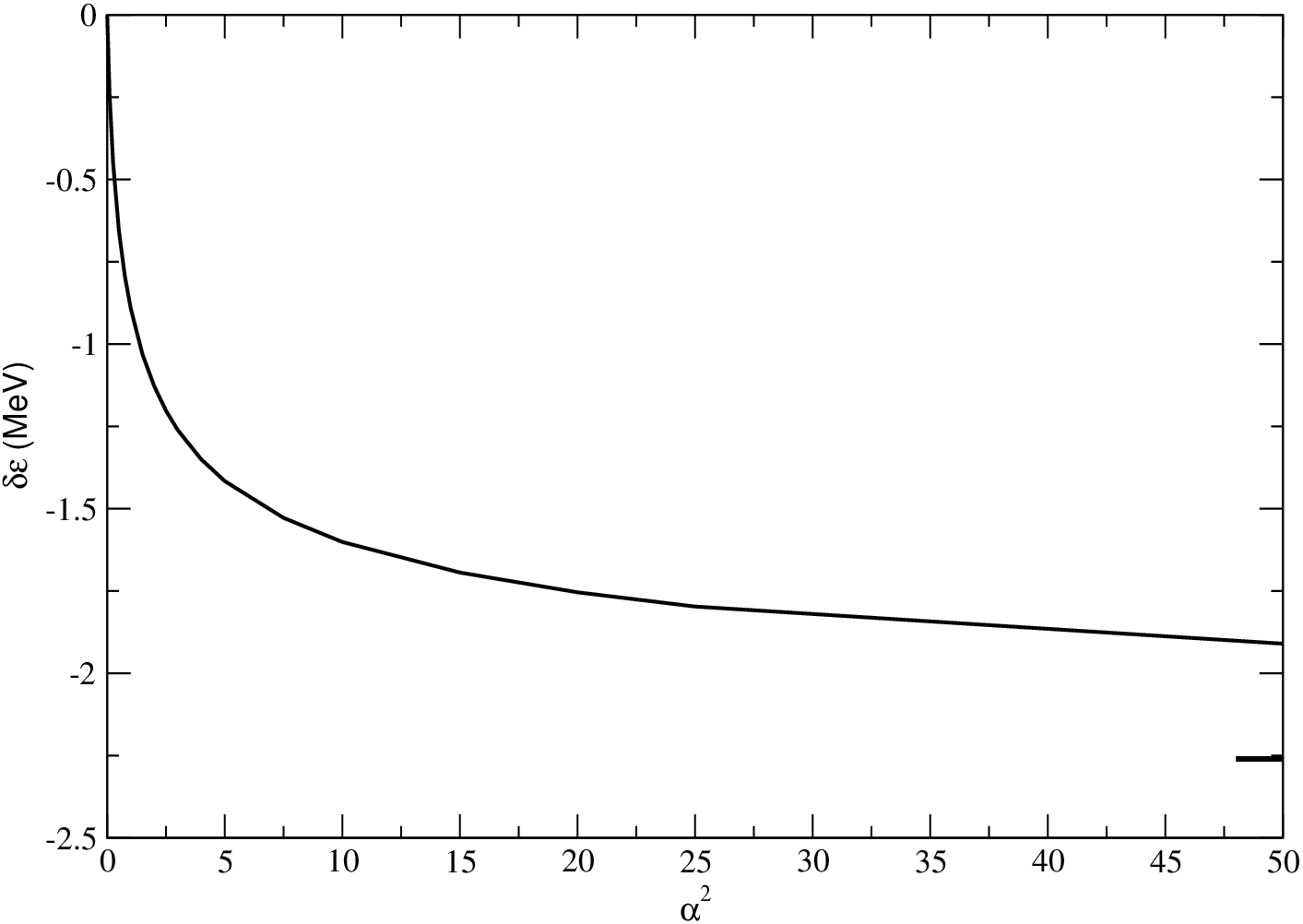}
\caption{Effect of the correlation function from Eq.\ (\ref{eqn:1}) on the total ground-state
splitting for $^{32}_{\Lambda}$S as a function of $\alpha^{2}$. The value of the splitting for
$\alpha^{2}=\infty$ is marked.}
\label{fig:corr3}
\end{center}
\end{figure}

The fact that the interactions take the form of Yukawa potentials here implies that there is some large 
contribution from short-distance physics that is influencing the calculation. To correct for this problem, 
a correlation function was introduced to remove the short-distance physics from the integrals. The following 
correlation function was used here
\begin{equation}
V_{corr}=\left(1-e^{-\alpha^{2}m_{V}^{2}(r_{>}-r_{<})^{2}}\right)
\label{eqn:1}
\end{equation}

\noindent A range of $\alpha^{2}$ was investigated for both $_{\Lambda}^{16}$O and $_{\Lambda}^{32}$S,
the effects of which are shown in Figs.\ (\ref{fig:corr1}) and (\ref{fig:corr3}) respectively. Note that, regardless of
$\alpha^{2}$, {\it the correlation function does not alter the level ordering of the doublet}; it changes only the 
magnitude of the splitting. Also, the cancellation that yields a small splitting in the ground-state of 
$_{\Lambda}^{16}$O is unaffected by the correlation function. Thus, we can improve the splittings
which were quite large while retaining the small splittings that resulted from cancellation.

Technically, the proper calculation in an effective field theory such as this one is to choose an appropriate 
cutoff, then add a contact term for each portion of the interaction, which are essentially just constants and 
can be fit to experiment. However, in this case not enough data is available for nuclei in the range of $A$ accessible
to this type of theory. The only relevant splitting that has been measured is the ground-state of $_{\Lambda}^{16}$O 
\cite{ref:Ta05,ref:Uk05}. Therefore, we conclude that at best one could claim to have a single contact term for all parts 
of the interaction. This is equivalent to a one parameter phenomenological calculation containing a correlation
function in coordinate space meant to simulate the proper calculation. 
\begin{table}
\begin{center}
\begin{tabular}{|c|c|c|c|c|c|c|} \hline\hline
Nucleus & $V(2)$ & $NT(1)$ & $\Lambda T(1)$ & $TT(1)$ & $TT(2)$ & $\delta\epsilon$ \\ \hline
$^{12}_{\Lambda}$B   & -26 & -3 & -7  & -68 & -17 & -122 \\ \hline
$^{16}_{\Lambda}$N   & -29 & 11 & -37 & 31  & 50  & 26   \\ \hline
                     & -20 & -2 & -6  & -58 & -14 & -100 \\ \hline
$^{16}_{\Lambda}$O   & -29 & 11 & -38 & 31  & 51  & 26   \\ \hline
$^{28}_{\Lambda}$Si  & -20 & -1 & -2  & -35 & -11 & -68  \\ \hline
$^{32}_{\Lambda}$S   & -15 & -7 & -21 & -68 & -13 & -124 \\ \hline
$^{40}_{\Lambda}$Ca  & -19 & 1  & -6  & 19  & 17  & 13   \\ \hline
                     & -22 & 1  & 4   & -24 & -5  & -47  \\ \hline
$^{48}_{\Lambda}$K   & -20 & 1  & -6  & 19  & 17  & 12   \\ \hline
$^{48}_{\Lambda}$Ca  & -10 & 0  & 0   & -16 & -11 & -38  \\ \hline
$^{88}_{\Lambda}$Rb  & -12 & 0  & -2  & 14  & 9   & 10   \\ \hline
$^{88}_{\Lambda}$Sr  & -6  & 0  & 0   & -8  & -5  & -19  \\ \hline
$^{208}_{\Lambda}$Pb & -1  & 0  & 0   & -3  & -2  & -7   \\ \hline
\end{tabular}
\caption{Calculation of the s$_{1/2}$-doublets (and some excited state splittings for $^{16}_{\Lambda}$N
and $^{40}_{\Lambda}$Ca) in select single $\Lambda$-hypernuclei (in keV) using the correlation function from Eq.\ (\ref{eqn:1}).
Here the value of $\alpha^{2}=0.044$ was used.}
\label{tab:split2}
\end{center}
\end{table}

The value of the cutoff that reproduced a splitting of +26 keV in the ground-state of $_{\Lambda}^{16}$O is 
$\alpha^{2}=0.044$; this translates into a momentum space cutoff of $\Lambda=\alpha^{2}m_{V}^{2} \sim 160$ MeV. The results of 
calculations conducted with this cutoff are shown in Table (\ref{tab:split2}). All of the splittings are now 
within the experimental error bars. 
\begin{figure}
\begin{center}
\includegraphics[scale=.40]{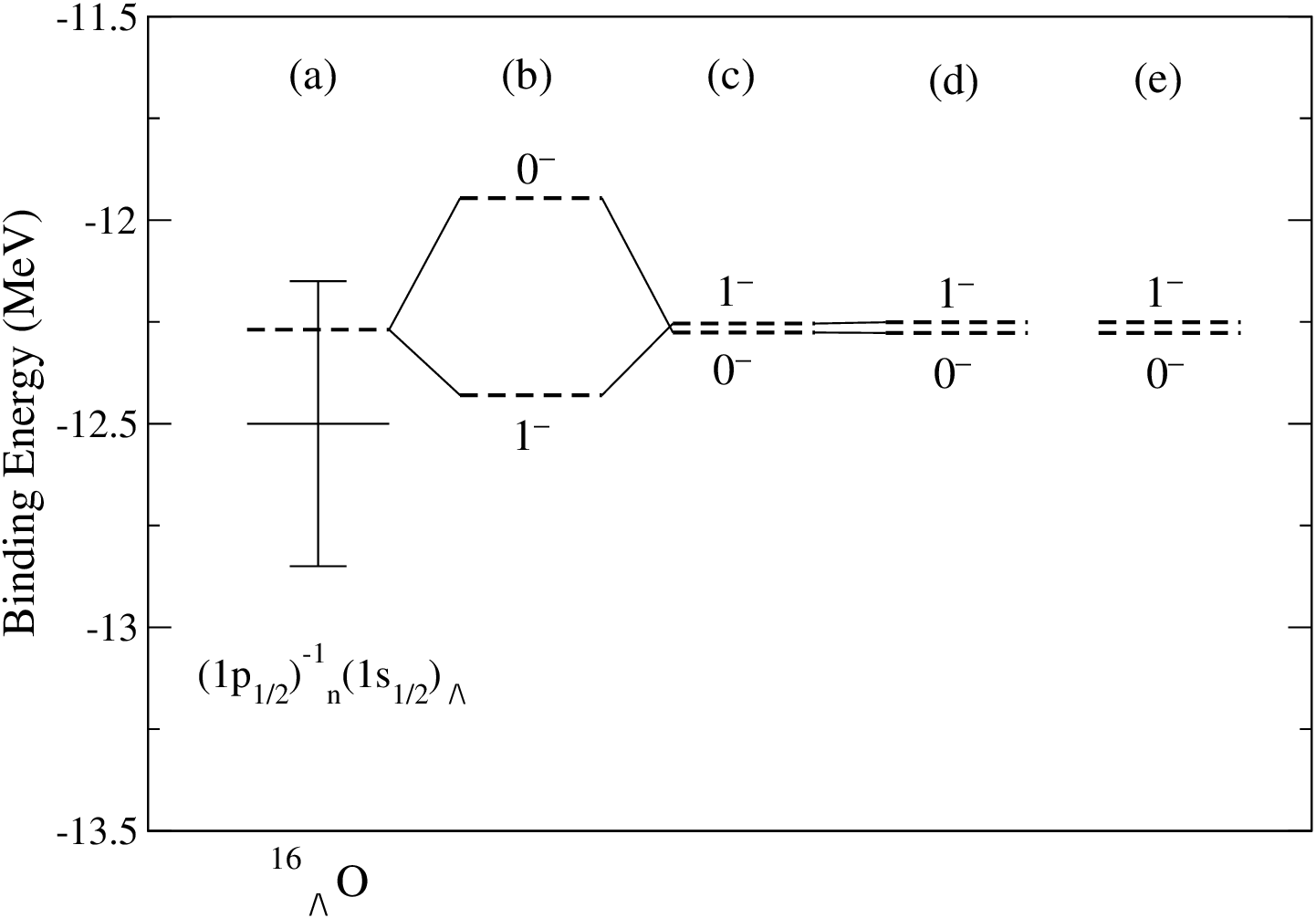}
\caption{The ground-state of $^{16}_{\Lambda}$O is plotted here. (a) is the experimental level and error bars 
\cite{ref:Pi91} along with the single-particle energy level \cite{ref:Mc04}, (b) is the splitting determined from $V(2)$, (c)
is the splitting determined from the expanded interaction in Eq.\ (\ref{eqn:TOT1}), (d) is the 
expanded interaction with the cutoff of $\alpha^{2}=0.044$, and (e) is the experimental splitting \cite{ref:Ta05,ref:Uk05}.}
\label{fig:split1}
\end{center}
\end{figure}

\begin{figure}
\begin{center}
\includegraphics[scale=.40]{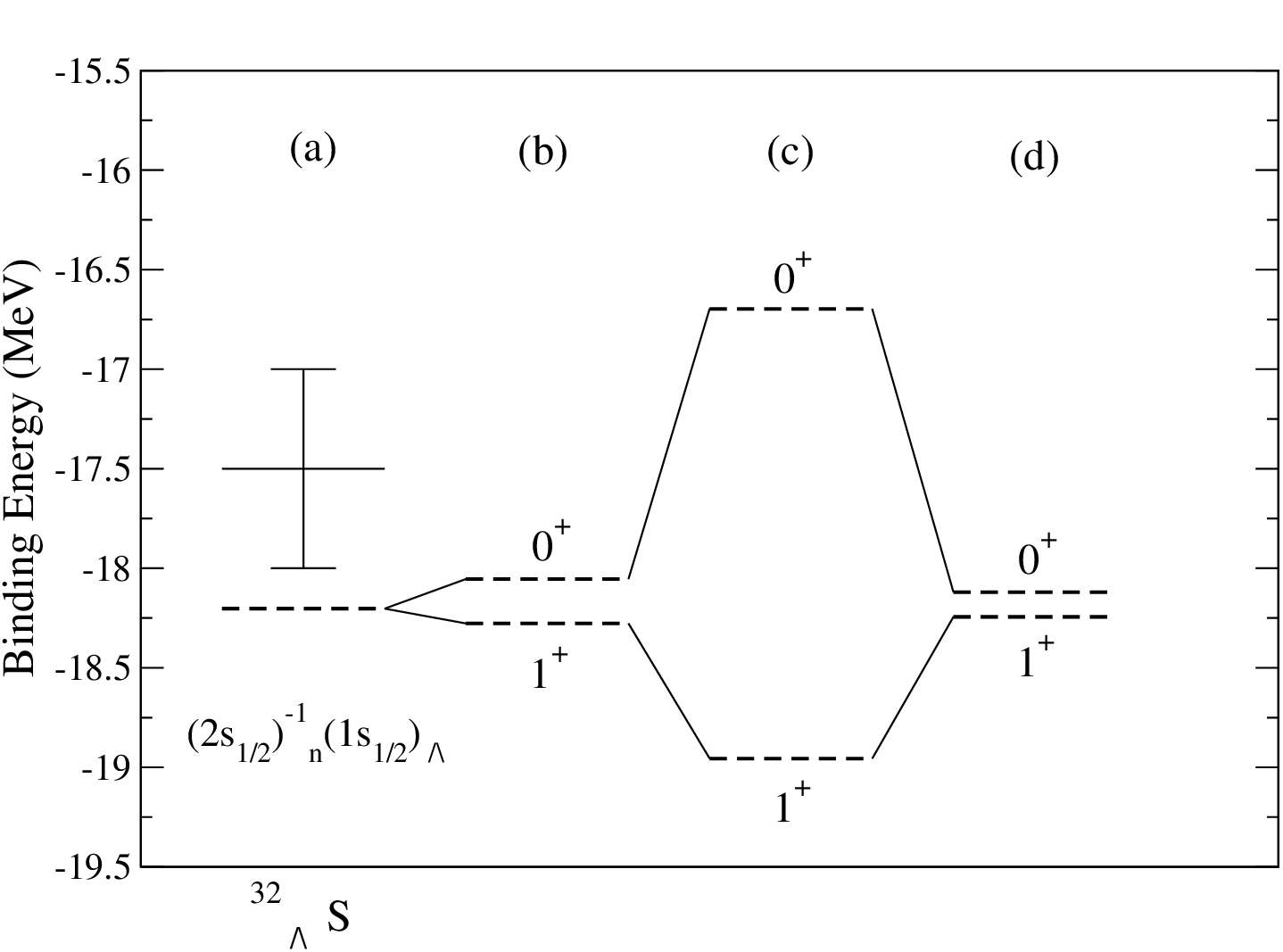}
\caption{The ground-state of $^{32}_{\Lambda}$S is plotted here. (a) is the experimental level and error bars 
\cite{ref:Be79} along with the single-particle energy level \cite{ref:Mc04}, (b) is the splitting determined from $V(2)$, (c)
is the splitting determined from the expanded interaction in Eq.\ (\ref{eqn:TOT1}), and (d) is the 
expanded interaction with the cutoff of $\alpha^{2}=0.044$.}
\label{fig:split2}
\end{center}
\end{figure}

\begin{figure}
\begin{center}
\includegraphics[scale=.40]{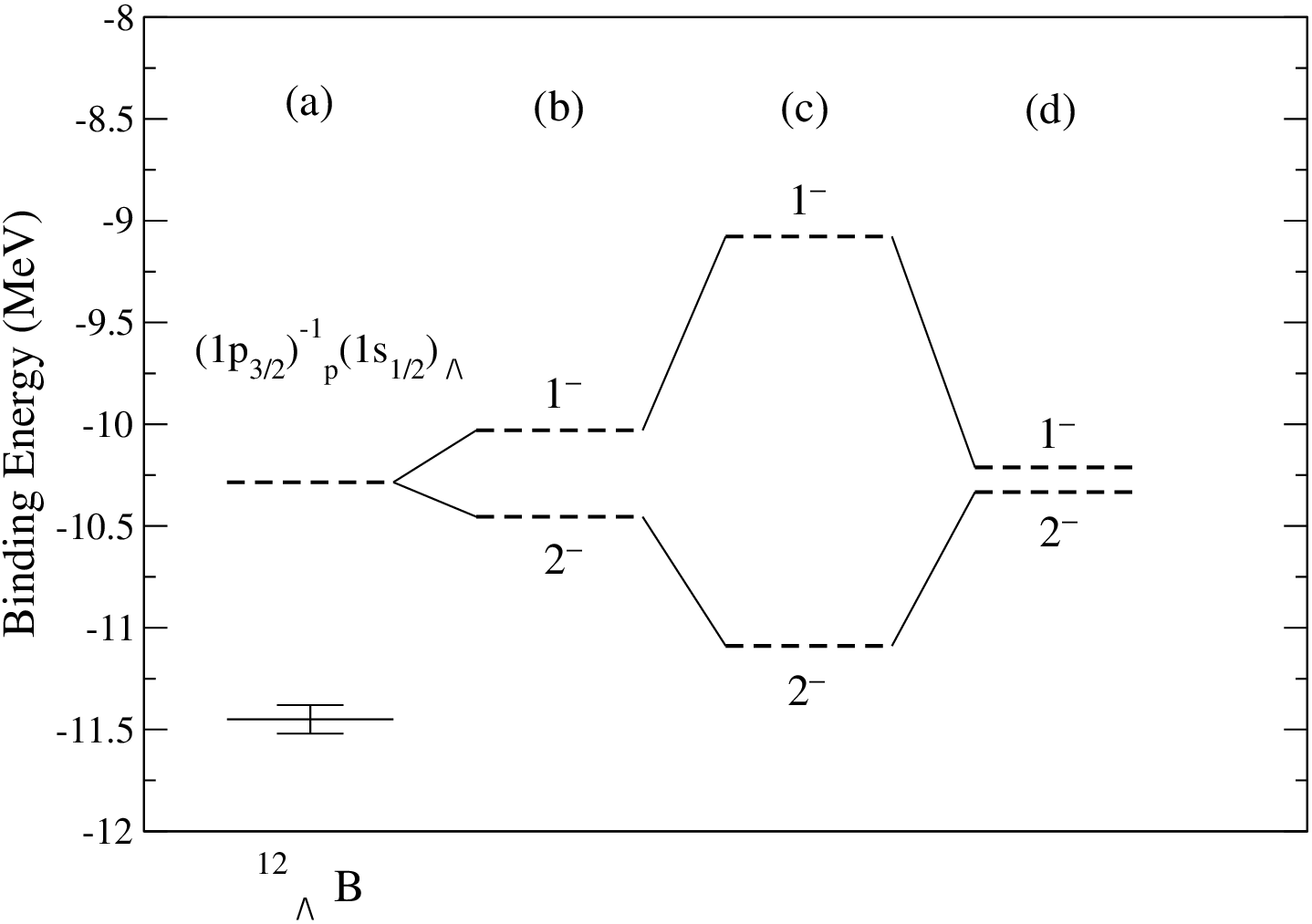}
\caption{The ground-state of $^{12}_{\Lambda}$B is plotted here. (a) is the experimental level and error bars 
\cite{ref:Ju73} along with the single-particle energy level \cite{ref:Mc04}, (b) is the splitting determined from $V(2)$, (c)
is the splitting determined from the expanded interaction in Eq.\ (\ref{eqn:TOT1}), and (d) is the 
expanded interaction with the cutoff of $\alpha^{2}=0.044$. \vspace{.4 in}}
\label{fig:split3}
\end{center}
\end{figure}

\begin{figure}
\begin{center}
\includegraphics[scale=.40]{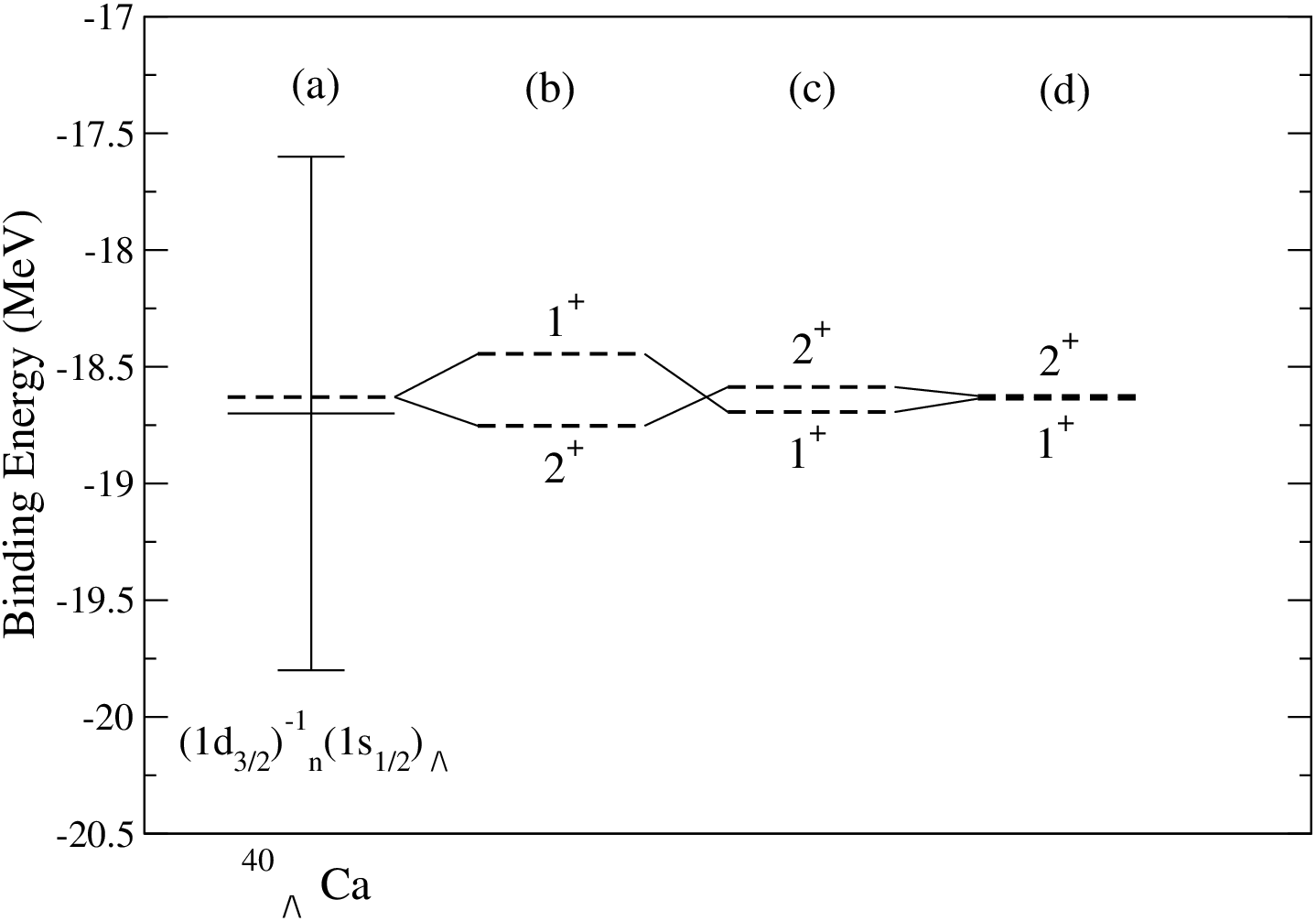}
\caption{The ground-state of $^{40}_{\Lambda}$Ca is plotted here. (a) is the experimental level and error bars
\cite{ref:Pi91} along with the single-particle energy level \cite{ref:Mc04}, (b) is the splitting determined from $V(2)$, (c)
is the splitting determined from the expanded interaction in Eq.\ (\ref{eqn:TOT1}), and (d) is the 
expanded interaction with the cutoff of $\alpha^{2}=0.044$.}
\label{fig:split4}
\end{center}
\end{figure}

\begin{figure}
\begin{center}
\includegraphics[scale=.40]{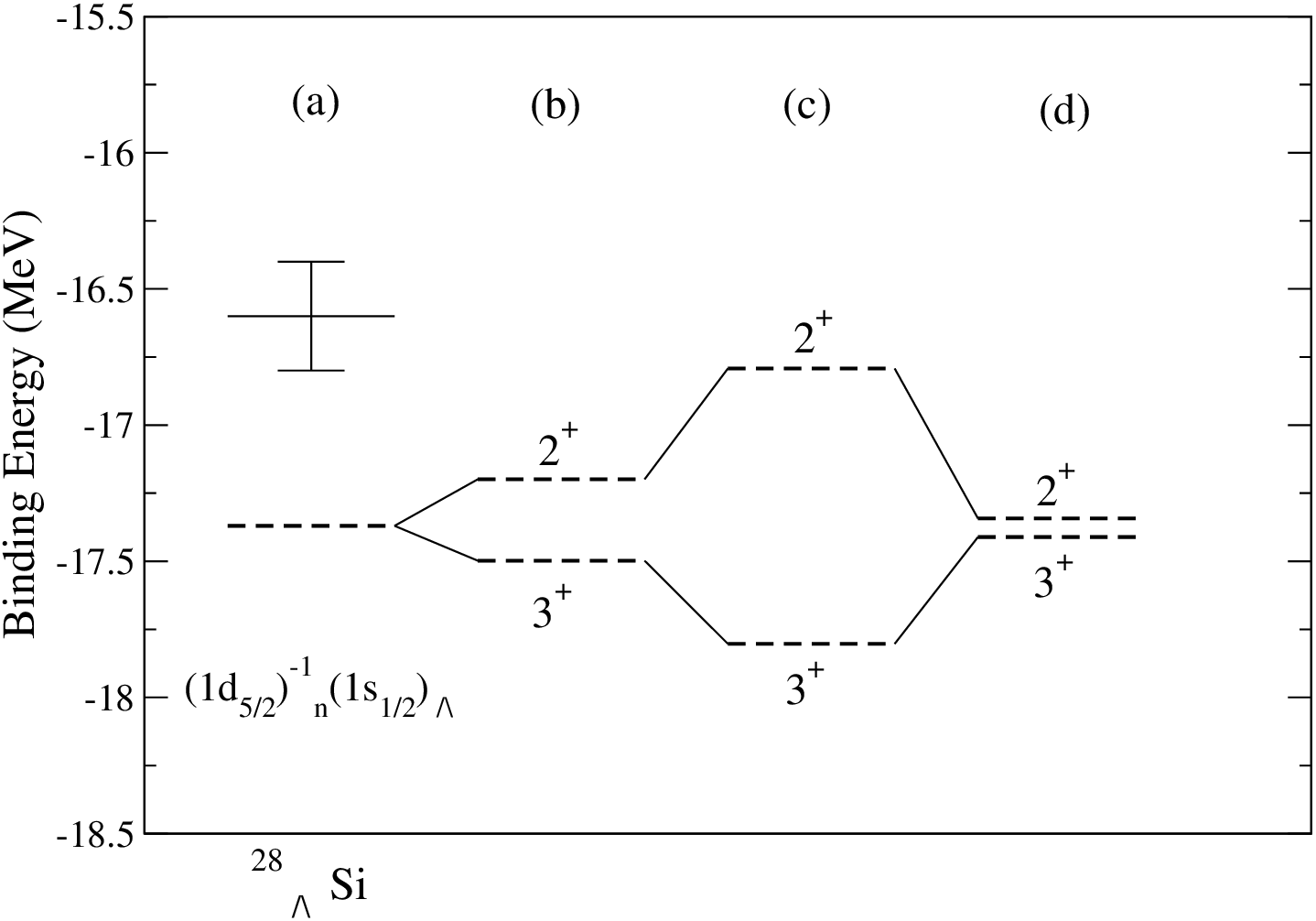}
\caption{The ground-state of $^{28}_{\Lambda}$Si is plotted here. (a) is the experimental level and error bars 
\cite{ref:Ha96} along with the single-particle energy level \cite{ref:Mc04}, (b) is the splitting determined from $V(2)$, (c)
is the splitting determined from the expanded interaction in Eq.\ (\ref{eqn:TOT1}), and (d) is the 
expanded interaction with the cutoff of $\alpha^{2}=0.044$.}
\label{fig:split5}
\end{center}
\end{figure}

Fig.\ (\ref{fig:split1}) shows the experimental error bars from the $\left(\pi^{+},K^{+}\right)$ reaction
\cite{ref:Pi91}, the single-particle energy level determined from the self-consistent equations, the splitting
and level ordering corresponding to the $V(2)$ contribution \cite{ref:Mc04}, the splitting and level ordering 
determined from the expanded interaction given by Eq.\ (\ref{eqn:TOT1}) without the correlation function and 
with the correlation function where $\alpha^{2}=0.044$, and the experimental doublet and level ordering \cite{ref:Ta05,ref:Uk05}
for $_{\Lambda}^{16}$O. Figs.\ (\ref{fig:split2}) -- (\ref{fig:split5}) show, for a range of nuclei taken from
Table (\ref{tab:split1}), the same contributions as Fig.\ (\ref{fig:split1}) except
that the experimental splittings have not been measured for these states. One can see from Figs.\ (\ref{fig:split1}) and
(\ref{fig:split4}) that the addition of the tensor force caused the level ordering to flip and decreased the
size of the splitting. In addition, the correlation function has only a limited effect on the size of the splitting. 
In contrast, one can see from Figs.\ (\ref{fig:split2}), 
(\ref{fig:split3}), and (\ref{fig:split5}) that the tensor force is an additive contribution to the spin-spin
force, hence the splitting becomes large. Also, the effect of the correlation function with this particular 
$\alpha$ is to decrease the splitting size to within the known experimental constraints. Again, note that
the level orderings for the expanded interaction are unaffected by the correlation function for all cases.

The predicted ground-state of $_{\Lambda}^{12}$B is $2^{-}$, as seen in Fig.\ (\ref{fig:split3}). This is inconsistent with
an analysis of the emulsion data in \cite{ref:Ki75} that determined the ground-state spin of this nucleus to be 1. However, 
this theory relies on spherical symmetry and there is evidence that $_{\Lambda}^{12}$B is heavily deformed.
Also, it should be mentioned that this nucleus may in fact be too small for this type of mean-field approach. 
A resolution to this discrepancy will be the subject of future work.

In \emph{conclusion}, we have developed a method to calculate the doublet splittings of select ground-state
single $\Lambda$-hypernuclei. This method consists of supplementing the self-consistent single-particle equations 
by constructing an effective interaction to simulate the residual particle-hole interaction. The form of the
effective interaction used here follows directly from the underlying lagrangian. Note that this formulation of 
the problem contains no free parameters. Retaining only the leading-order interaction terms, this calculation was
conducted in \cite{ref:Mc04}; this level of truncation in the residual interaction was inadequate to describe
either the doublet size or level ordering in the ground-state of $_{\Lambda}^{16}$O. To improve on this calculation,
we included in this effective interaction the contributions that contained gradient couplings to the neutral vector field; 
this incorporated a tensor force into the calculation known to play a crucial role in these systems that did not 
appear at leading-order. Cancellation occurs for the states that satisfy $j_{1}+j_{2}+\pi=even$, flipping the sign from 
the simple leading-order spin-spin interaction. However, the contributions are additive for the states satisfying 
$j_{1}+j_{2}+\pi=odd$, resulting in splittings that lie outside the known experimental error bars. It turns out that 
the integrals are dominated by short-distance physics; as a result, a cutoff was introduced to reduce this contribution. 
This cutoff did not effect the level orderings of any state; it did however, reduce the size of the splittings for 
states with $j_{1}+j_{2}+\pi=odd$ to within the experimental constraints while simultaneously retaining the 
cancellation the yielded small splittings in the states $j_{1}+j_{2}+\pi=even$. Thus, we obtain a realistic
description of the effect of the tensor couplings on the doublet orderings and splittings. 

I would like to thank Dr. B. D. Serot and Dr. J. D. Walecka for their support and advice, their careful reading of
the manuscript, and their helpful comments. This work is supported by Indiana University and the State of Indiana at
the Indiana University Nuclear Theory Center, 2401 Milo Sampson Lane, Bloomington, IN, 47408.

% \appendix

% \section{}
% \label{}

\end{document}